\documentclass[journal=jctcce,manuscript=article]{achemso}
\setkeys{acs}{usetitle=true}
 
\usepackage{graphicx}
\usepackage{dcolumn}
\usepackage{multirow} 
\usepackage{bm}
\usepackage{blindtext} 
\usepackage{tabularx} 
\usepackage{tikz}

\newcolumntype{L}[1]{>{\raggedright\arraybackslash}p{#1}}
\newcolumntype{C}[1]{>{\centering\arraybackslash}p{#1}}
\newcolumntype{R}[1]{>{\raggedleft\arraybackslash}p{#1}}

\author{Simone Sturniolo}
\email{simone.sturniolo@stfc.ac.uk}
\author{Leandro Liborio}%
\author{Samuel Jackson}
 \affiliation{Scientific Computing Department, UKRI}

\title{Comparison between Density Functional Theory and Density Functional Tight Binding approaches for finding the muon stopping site in organic molecular crystals}

\date{\today}

\keywords{Muon spectroscopy, random structure search, DFT, tight binding, software}

\begin{document}

\begin{abstract}
Finding the possible stopping sites for muons inside a crystalline sample is a key problem of muon spectroscopy. In a previous work, we suggested a computational approach to this problem, using Density Functional Theory software in combination with a random structure searching approach using a Poisson sphere distribution.\newline
In this work we test this methodology further by applying it to three organic molecular crystals model systems: durene, bithiophene, and tetracyanoquinodimethane (TCNQ). Using the same sets of random structures we compare the performance of Density Functional Theory software CASTEP and the much faster lower level approximation of Density Functional Tight Binding provided by DFTB+, combined with the use of the \texttt{3ob-3-1} parameter set. We show the benefits and limitations of such an approach and we propose the use of DFTB+ as a viable alternative to more cumbersome simulations for routine site-finding in organic materials. Finally, we introduce the Muon Spectroscopy Computational Project software suite, a library of Python tools meant to make these methods standardized and easy to use.
\end{abstract}

 if keyword
\maketitle

\section{\label{sec:intro} Introduction}

Muon spin spectroscopy is a technique in which a beam of polarized positive muons is implanted in a solid sample, often with the additional application of an external magnetic field, and the pattern of positron emission caused by their decay is then observed and used to measure the properties of the material. Since the positron emission depends on the direction of the muon's spin at the time of its decay, observation of its time evolution can allow one to infer information about the magnetic structure of the sample thanks to the hyperfine interaction between electrons and the muon.\newline
A key step of understanding muon spectra is to predict where in a crystal would an incoming muon effectively come to rest. This stopping site problem is well known and has been tackled in a number of different ways \cite{Moller:2013}, including by use of simulations of various levels of accuracy. In our previous work \cite{Liborio:2018} we proposed a new method to approach this problem, by combining the Ab-Initio Random Structure Search methodology, or AIRSS \cite{Morris:2008,Morris:2009} with a new random generation technique making use of a Poisson sphere distribution \citep{Lagae:2006} and the clustering techniques implemented in the Python library Soprano \cite{soprano}.\newline
The main performance cost in the proposed approach consisted in the use of Density Functional Theory (DFT) software to perform the necessary geometry optimization calculations. In this paper, we compare the results obtained with that approach to those obtained with a higher level approximation, Density Functional Tight Binding (DFTB) for a specific class of materials, namely, organic molecular crystals. Our proposal is that DFTB can be used as a computationally cheaper substitute for DFT, thus making calculations for systems under one hundred atoms easily feasible in a few hours on a single desktop machine, and opening the way to the study of larger systems like proteins and polymers. We also introduce new clustering techniques more suitable to treating systems with a large amount of possible stopping sites.

\section{\label{sec:bckg}Background}

While muon spectroscopy is often used to study inorganic systems with peculiar magnetic properties such as high temperature superconductors, there is an interest in using them in organic materials as well \cite{Percival:1987,Dediu:2009}. When a positive muon is implanted in a material it can either stop as is or capture an electron and form a bound state that is known as "muonium". Muonium acts effectively like a lighter hydrogen atom, meaning it possesses the same chemical properties and tends to react in the same ways. The main differences in behavior concern its dynamical properties and the entity of quantum delocalization effects it undergoes - both things dependent on its mass. In this paper, we focus on equilibrium stopping sites of muonium treated classically, and thus both these differences are irrelevant, and muonium can be treated as if it was a hydrogen atom.\newline
In our previous work, we exploited this similarity to treat crystalline systems with an added muon with the DFT software package CASTEP. The muon, in these calculations, was simply represented by a labeled hydrogen atom. The core idea was to use the AIRSS methodology - create a large number of copies of the same host structure in each of which the muon occupied a different starting position and then carrying out a geometry optimization to find the closest energy minimum. This method is justified by the observation that in general the energy landscape for atomistic systems seems to be structured in such a way that the attraction basins for each energy minimum have a size that scales with the depth of that minimum \cite{Doye:1998,Doye:2005,Massen:2007}. Therefore, by starting with random configurations, one has a high likelihood to find the lowest energy minima.\newline
If we consider the nature of this problem, it is not unreasonable to think that it could be acceptable to replace DFT with a cheaper computational method, as long as some key properties were guaranteed. While a new method could describe an overall different energy landscape for the same system, one only needs it to satisfy two conditions for it to be a viable replacement for DFT in this scenario:

\begin{enumerate}
\item that the resulting landscape has at least as many or more energy minima as the DFT landscape;
\item that for each minimum in the DFT landscape there exists at least one minimum in the new landscape that falls close to it, or at least within the same attraction basin as defined in the DFT landscape.
\end{enumerate}

Given these two conditions, it is easy to see how one can gain much by replacing DFT with the new, cheaper method. If, for example, 100 random structures are generated, one could find that after optimization they only result in, say, 10 minima. These 10 minima can then be further optimized with DFT, leading to the same results that we would have obtained from a pure AIRSS+DFT calculation. If the new method is cheap enough that its computational cost is trivial with respect to that of a DFT calculation, this results effectively in an almost tenfold increase in speed compared with optimizing all the original 100 structures with DFT.\newline
In this paper, we propose that, at least for organic molecular crystals, Density Functional Tight Binding (DFTB) \cite{Andersen:1984,Seifert:2007} might be such a method. To validate this idea we compare the muon stopping sites found with both DFT and DFTB starting from the same random generated configurations for three organic molecular crystals.

DFTB is an electronic structure method that, much like DFT, makes use of the Kohn-Sham approximation to solve the quantum many body problem for electrons. It is however approximate in that it only represents these electrons as bound to the ions, and does not describe the full spatial wavefunction. This allows it to be computationally cheaper than a DFT calculation. It is also not ab initio, because it makes use of parametrizations computed from pure DFT calculations to describe interactions between chemical species \cite{Porezag:1995}. These parametrizations are stored in sets of so-called Slater-Koster files which are computed to cover only specific groups of elements, and thus can not treat any other species. The choice of organic molecular crystals was in fact prompted also by the availability of a well-documented Slater-Koster parameter set for organic compounds in the \texttt{3ob-3-1} set \cite{Gaus2013,Gaus2014}, that covers among the other elements carbon, hydrogen, oxygen, nitrogen and sulphur. Success with this specific subset of chemistry could encourage the creation of more system-specific parametrizations in the future to treat other classes of compounds.

\begin{figure}
\includegraphics[]{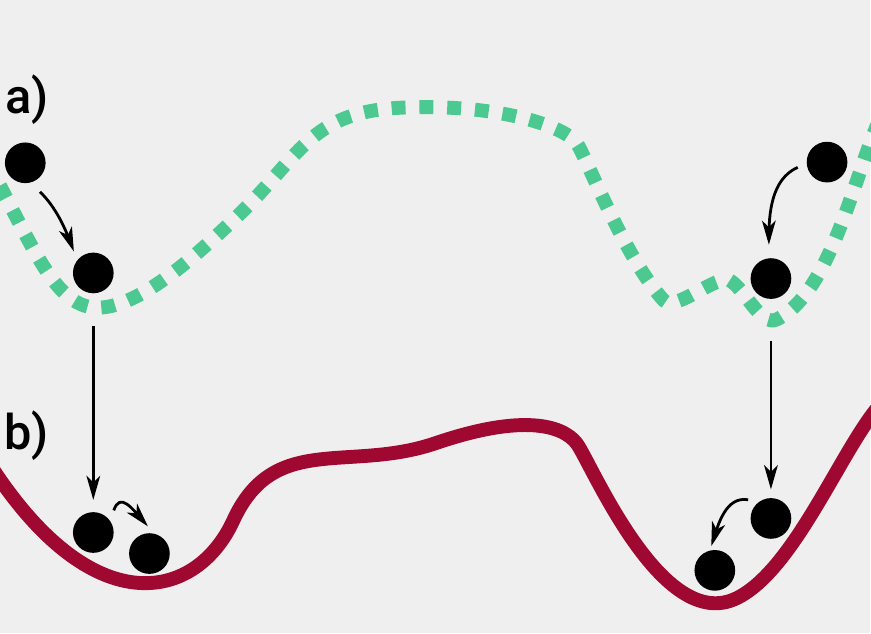}
\caption{An illustration of the correspondence between two similar but not identical energy landscapes for the same system. Landscape a) represents the more high-level, approximate theory, whereas landscape b) represents the more fundamental one. Landscape a) has minima that only roughly correspond to the ones in b); in some cases, more than one of the minima of a) map to the same minimum of b). However two particles that start with an optimization in a) will eventually find the minima of b) if optimized further.}
\label{fig:landscape}
\end{figure}

\section{\label{sec:calc}Calculations}

\subsection{\label{subs:mols}Choice of molecules}

In this paper, we focused on three exemplar, crystalline, molecular-organic systems:  tetracyanoquinodimethane (TCNQ), bithiophene and durene. These systems represent the class of organic crystals that are most studied using muon techniques while, at the same time, being of enough individual practical interest.

\begin{figure}[h]
\includegraphics[scale=0.7]{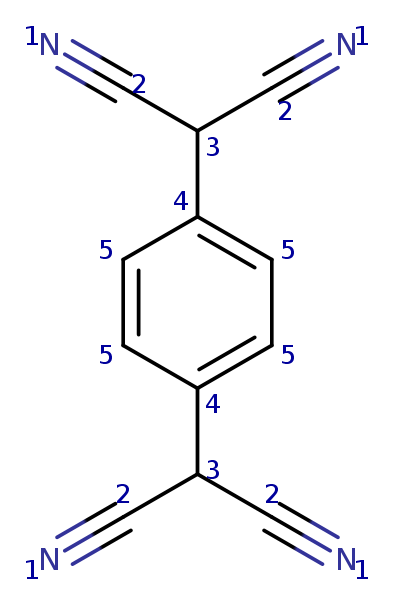}
\caption{Structural formula of TCNQ with numbered molecular sites. Equivalent sites have the same number.}
\label{fig:TCNQ-numbered}
\end{figure}

\begin{figure}[h]
\includegraphics[scale=0.4]{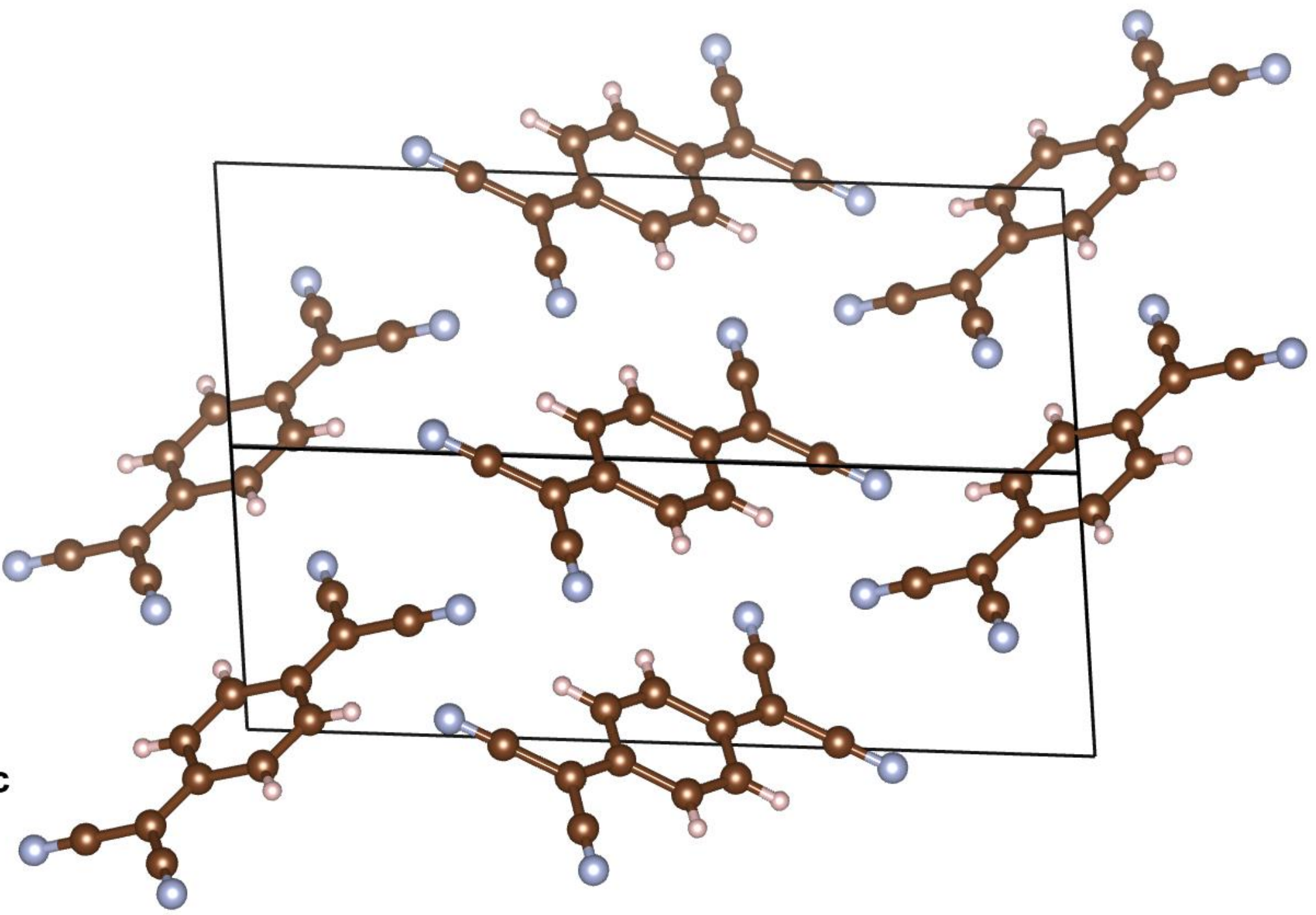}
\caption{Crystalline structure for TCNQ. The view is along the [110] direction and the individual TCNQ molecules are easily visualized. C atoms are brown, H atoms beige and N atoms metallic blue.}
\label{TCNQ-fig}
\end{figure}

TCNQ is a cyanocarbon with formula $\mathrm{(NC)_2CC_6H_4C(CN)_2}$ (Figure \ref{fig:TCNQ-numbered}). It crystallises around a temperature of 566 K in a structure with four molecules per unit cell (Z=4) and space group C2/c (Figure \ref{TCNQ-fig}).  It can form charge-transfer salts that have relatively large electrical conductivity. For instance, when TCNQ is combined with the electron donor tetrathiafulvene (TTF), it forms the TTF-TCNQ complex, in which TCNQ is the electron acceptor \cite{torrance:1975}.  Muon experiments were performed on this compound \cite{pratt:2000} to investigate its electronic structure; a key step of interpreting them is then to determine the relationship between the muon radical states and the electronic states that are formed in the muoniated molecule.

\begin{figure}[h]
\includegraphics[scale=0.7]{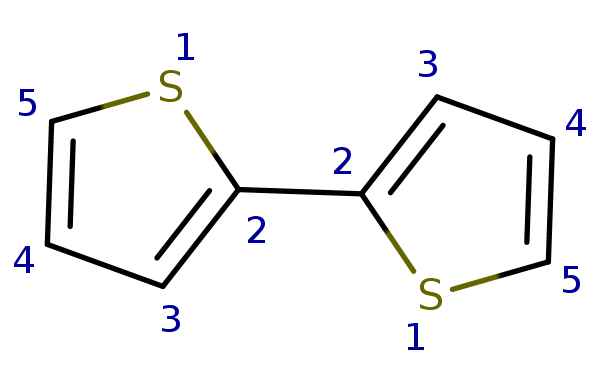}
\caption{Structural formula of bithiophene with numbered molecular sites. Equivalent sites have the same number.}
\label{fig:bithiophene-numbered}
\end{figure}

\begin{figure}[h]
\includegraphics[scale=0.5]{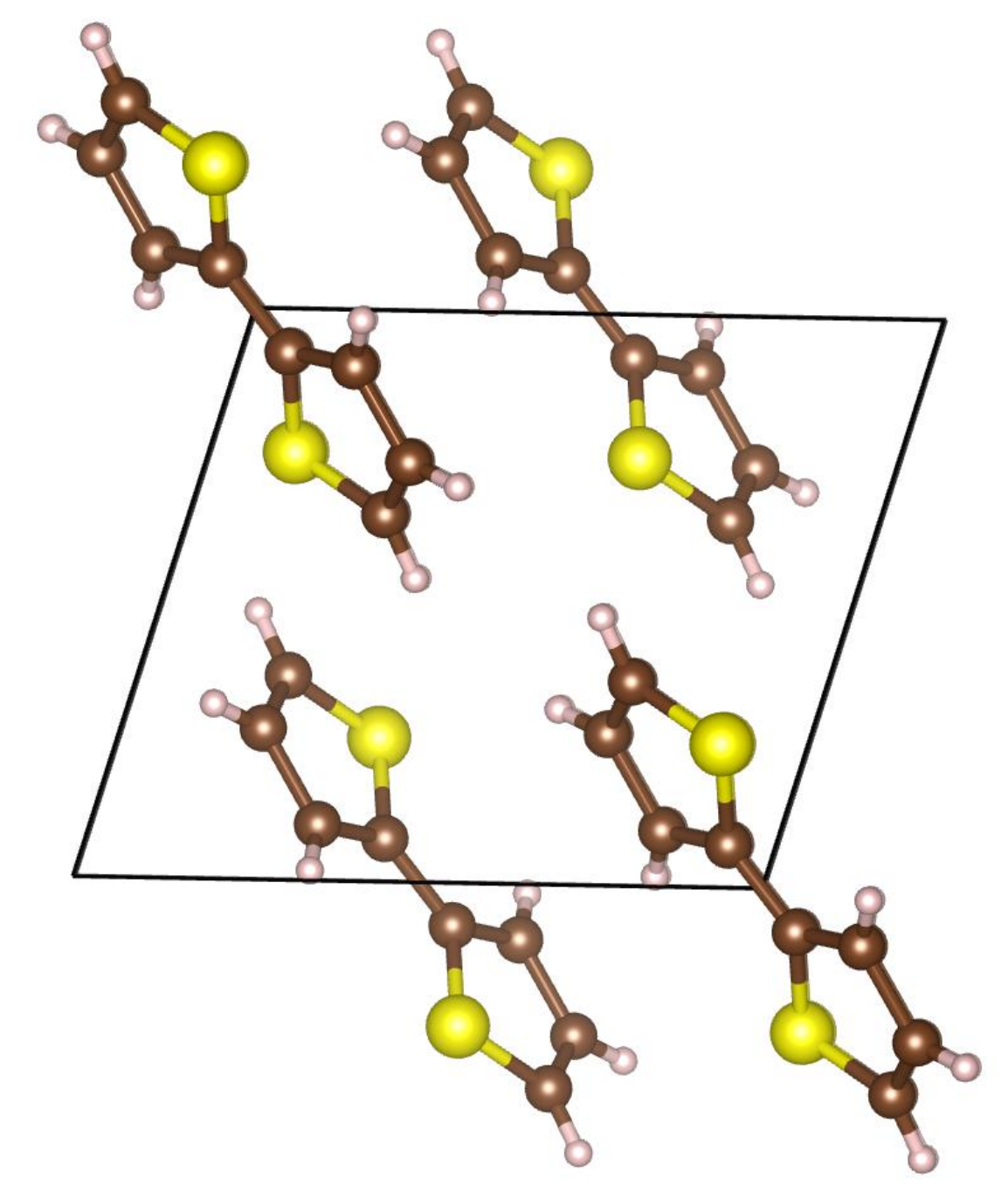}
\caption{Bithiophene's crystalline structure. The view is along the [010] direction and the individual molecules are easily visualized. C atoms are brown, H atoms beige and S atoms yellow.}
\label{bithiophene-fig}
\end{figure}

2-2'-bithiophene is the dimer of thiophene, an heterocyclic compound whose polymers are of great interest due to their conductive properties. Bithiophene has formula $\mathrm{(C_4H_3S)_2}$ (Figure \ref{fig:bithiophene-numbered}) and crystallizes at 306 K in a structure with two molecules per unit cell (Z=2) and space group P2\_1/c (Figure \ref{bithiophene-fig}). Using the electrochemical polymerization method, bithiophene is used as a precursor for polythiophenes (PTs) more often than thiophene itself\cite{INGANAS:1985}, and muons have been used to study the microscopic charge transport processes in polythiopene-based polymers. In particular, muons have been successful in explaining the 1-dimensional intra-chain charge transport mechanisms and 3-dimensional inter-chain charge hopping effects that are related to the electrical and optical properties of these polythiopene-based polymers\cite{RISDIANA:2012}.  

\begin{figure}[h]
\includegraphics[scale=0.4]{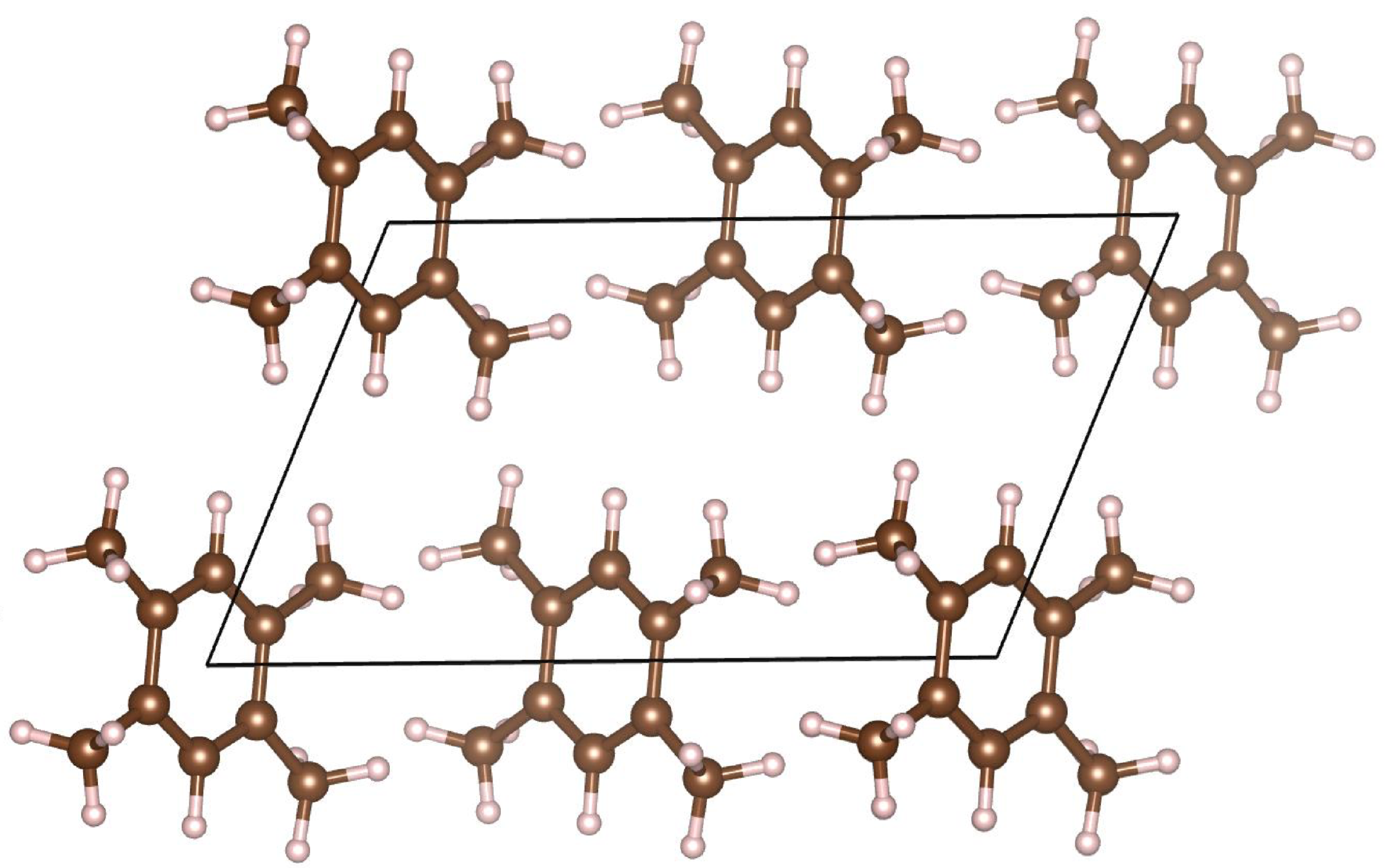}
\caption{Crystalline structure for durene. The view is along the [010] direction and the individual durene molecules are easily visualized. C atoms are brown and H atoms beige.}
\label{durene-fig}
\end{figure}

Finally, durene (1,2,4,5 tetramethyl-benzene) has formula $\mathrm{C_6H_2(CH_3)_4}$ and crystallizes at 352.3 K in a structure with two molecules per unit cell (Z=2) and space group P2\_1/c (Figure \ref{durene-fig}). It was one of the first substituted aromatic compounds to be studied with muons, and two muon stopping sites were suggested, located at the C-H sites in the ring, from above and below molecular plane \cite{RODUNER:1981}. Aromatic compounds with different substituting groups such as durene are a useful model to study classical and quantum dynamics of muons in molecular solids. Combined experimental and theoretical studies on crystalline benzene showed the effect of classical dynamics, but no final evidence of quantum tunneling  was observed \cite{sturniolo:2017}. Comparison to similar results in a family of selectively substituted benzene-based crystalline solids, with modified dynamics due to mass and steric effects, can help elucidate the issue. 

In all these materials, the main effects of muon implantation are the geometrical distortion of the surroundings of the muon stopping site and the presence of an unpaired electron in the crystalline structure. The unpaired electron in these muoniated compounds goes to the highest un-occupied molecular state. If the extent of the distortion is small, the newly occupied molecular state in the muoniated compounds can be used to study the occupied electronic states that result from charge transfer.  Hence, knowledge about the muon stopping sites in these materials, as well as about the parameters of the hyperfine interactions between the muon and the molecular electronic states, can be very helpful. The main structural characteristics of TCNQ, bithiophene and durene are summarized in Table \ref{tab:molprops}.

\begin{table*}[ht]
\begin{center}
\scalebox{0.8}{\begin{tabular}{ |c|c|c|c|c|c|} 
 \hline
 \textbf{Compound} & \textbf{Formula} &\textbf{Mol. per cell} & \textbf{Space Group} & \textbf{Lattice Parameters [$\AA$]} & \textbf{Angles }[Degrees] \\
 \hline
 bithiophene & $C_{8}H_{6}S_{2}$ & 2 & P 21/c &  (7.734, 5.729, 8.933) & (90.00, 106.72, 90.00) \\ 
 durene & $C_{10}H_{14}$ & 2 & P 21/a & (11.590, 5.740, 7.040) & (90.00, 112.80,  90.00)  \\ 
 TCNQ & $C_{12}H_{4}N_{4}$ & 4 & C 2/c & (8.896,  6.913, 16.439) &  (90.00,  98.29,  90.00) \\ 
 \hline
\end{tabular}}
\end{center}
\caption{Structural characteristics of bithiophene, durene and TCNQ.}
\label{tab:molprops}
\end{table*}

\begin{figure}[h]
\includegraphics[scale=0.7]{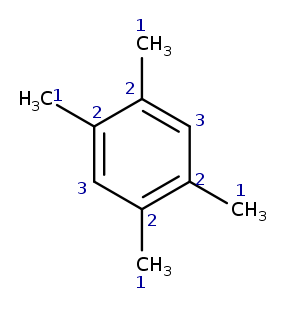}
\caption{Structural formula of durene with numbered molecular sites. Equivalent sites have the same number.}
\label{fig:durene-numbered}
\end{figure}

\subsection{\label{subs:pars}Geometry Optimization}

Random starting configurations have been generated using the Poisson sphere algorithm originally described in a previous work \cite{Liborio:2018}. This algorithm guarantees that all starting positions for the muon are at least at a distance $r_{min}$, here chosen to be 0.8 \AA{}. Additionally, a constraint was enforced to keep the generated muons at a distance from the ions equal to:

\begin{equation}\label{eq:vdw}
	r_{min}^{(ion)} = s\frac{r_{vdw}^{(Mu)}+r_{vdw}^{(ion)}}{2}
\end{equation}

where $r_{vdw}$ are the Van der Waals radii used for the given elements (in this context, muonium is treated as hydrogen) and $s$ is a scaling factor that in this case was set to $s=0.5$. The values of the used Van der Waals radii for various elements are reported in Table \ref{tab:vdw}.

\begin{table}
\begin{tabular}{|c|c|c|c|c|}
\hline
Element & H & C & N  & S\\
\hline
$r_{vdw}$ (\AA{}) & 1.2 & 1.95 & 1.55 & 2 \\
\hline
\end{tabular}
\caption{Van der Waals radii for elements appearing in the tested molecules.}
\label{tab:vdw}
\end{table}

The generation process continues until the entire free available space is filled and any further attempt to add another muon fails to meet the constraints. The process generated 396 different starting structures for bithiophene, 448 for durene, and 529 for TCNQ.\newline
Geometry optimization on these structures was then performed with a BFGS algorithm, with fixed unit cell parameters, and using both CASTEP 8.0 and DFTB+ 17.1, to a tolerance of 0.05 eV/\AA{} for the forces. For CASTEP \cite{Clark:2005}, the PBE exchange-correlation functional was used, in combination with a standard Tkatchenko-Scheffler scheme \cite{McNellis:2008,Tkatchenko:2009} for dispersion forces and auto-generated ultrasoft pseudopotentials. \newline 
DFTB+\cite{Aradi:2007} is a code that implements the Density Functional Tight Binding algorithm plus additional features, such as dispersion forces corrections. For these calculations, we made use of the self-consistent charges scheme \cite{Elstner:1998} and 3rd order corrections \cite{Yang:2007}. As mentioned already, the Slater-Koster files were those of the \texttt{3ob-3-1} parametrization \cite{Gaus2013,Gaus2014}. We also experimented with the use of dispersion correction \cite{Elstner:2001}, but it caused convergence to become extremely slow, possibly because of the small size of the unit cell used. Since our main purpose was to test an alternative to CASTEP that would be as quick and cheap as possible -while still producing similar results- we decided not to use dispersion corrections in DFTB+ calculations.\newline
The plane wave cutoff for CASTEP as well as the size of the k-point grid for both CASTEP and DFTB+ were chosen by converging energy and forces to a tolerance inferior to that required by the calculation for each individual system. Values of 700 eV for the CASTEP plane wave cutoff and 2x2x2 for either k-point grid were found satisfying and used for all systems.

\subsection{\label{subs:clust}Clustering}

\begin{figure}
\includegraphics[width=0.5\textwidth]{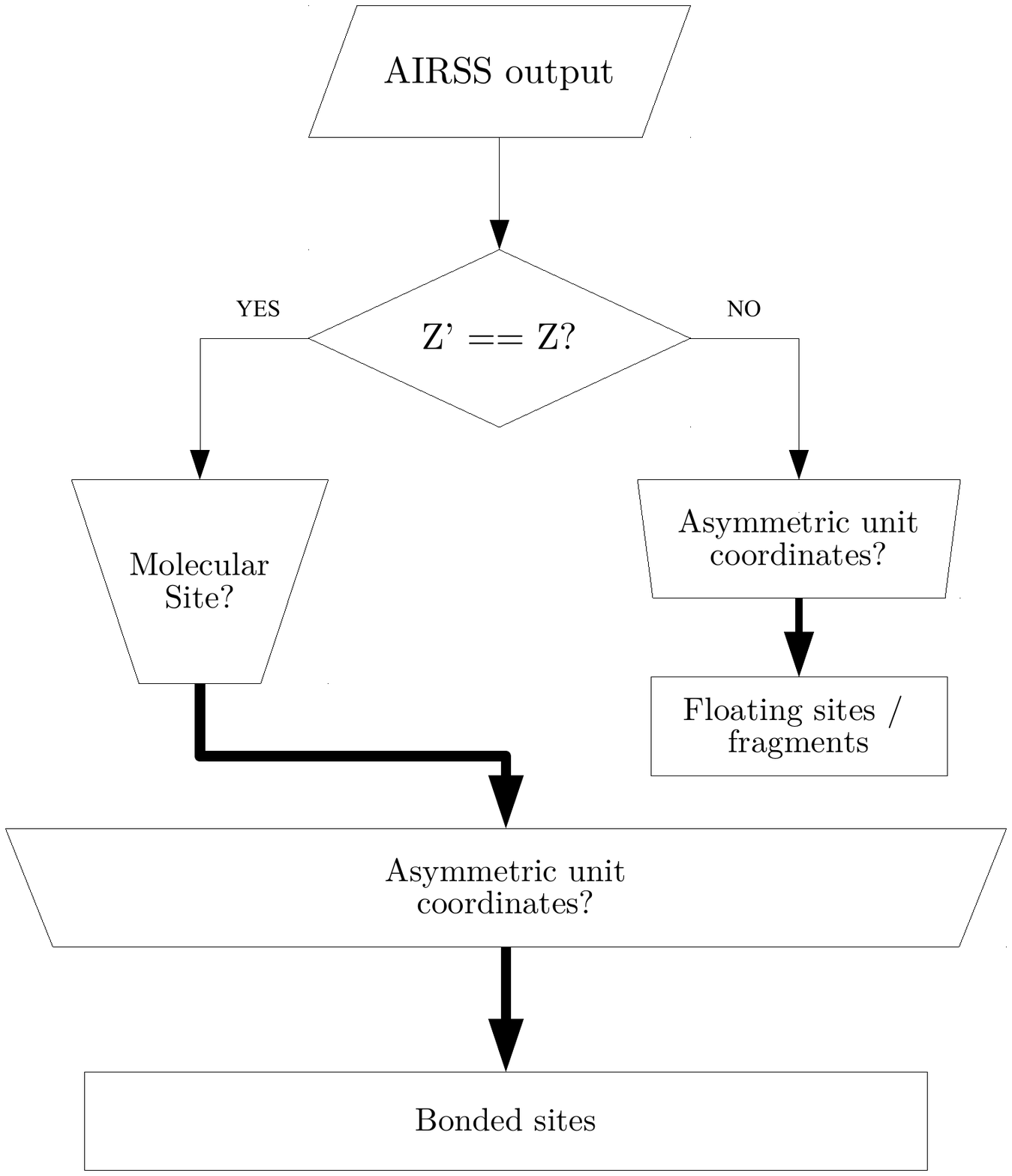}
\caption{Workflow for the clustering process. $Z$ and $Z'$ represent respectively the number of molecules in the pure structure and the optimised structure with an added muon as computed by Soprano. Thick arrows represent already partially clustered structures.}
\label{fig:workflow}
\end{figure}

In a previous work \cite{Liborio:2018} we already showed how random structure searching could be used fruitfully to identify muon sites in some simple semiconductor crystals. In that work, the optimized structures were analyzed using the software library Soprano \cite{soprano} and applying basic clustering techniques.\newline
In this paper we expand on that methodology by establishing a work flow of successive classification processes based on discrete variables or the clustering of continuous ones. The main reason for this is that the number of predicted sites for the molecular crystals is much higher than anything seen in the simple crystals treated before. While a single clustering step can indeed be used, meaningful interpretation of the data with that approach becomes much harder. It was therefore deemed more appropriate to use this multi-step process in order to highlight the physically meaningful differences between groups of clusters. The full workflow is illustrated in Figure \ref{fig:workflow}. The key steps are as following:

\begin{enumerate}
\item the number of individual organic molecules in the optimized structure with added muonium, $Z'$, is computed by making use of Van der Waals radii (methods included in Soprano) and compared to the number of molecules in the structure without defects, $Z$. If $Z$ and $Z'$ are different, this is a sign that: \newline
a) the muon has not bonded with any molecules (in which case $Z' = Z+1$) and it is sitting in an intermolecular equilibrium configuration that we termed a 'floating site'; or \newline
b) that the muon has acted as a bridge joining two molecules (in which case $Z' = Z - 1$), or that some other process by which, for example, molecular fragments have detached has taken place. Of these processes, the 'floating site' one is the only one observed consistently through both CASTEP and DFTB+ simulations in significant numbers of structures. The three groups ($Z' = Z$, $Z' = Z+1$, and every other case) are separated, and the latter two are subjected directly to step 3.
\item in structures in which $Z'= Z$, the muon has bonded with one of the existing molecules. For these, the molecule containing the muon is considered and the bonding site is identified automatically analysing the bonding structure, using a method suitable to each system and the library NetworkX for network analysis \cite{networkx}. For TCNQ and bithiophene, the network distance from the muon and the two nitrogen and sulphur atoms respectively was considered. For durene, it was the bonding structure of the nearest carbon. The details of these methods and how they identify the main sites are explained in the Supplementary Information (S1).  Figures (\ref{fig:TCNQ-numbered}), (\ref{fig:bithiophene-numbered}) and (\ref{fig:durene-numbered}) show the numbered molecular sites where the muon can bond to the different molecules while maintaining the structural integrity of the molecules. There are, however, particular cases where the muon bonds and breaks the structural integrity of the molecule, i.e.: breaks an aromatic ring. In these cases, the identification string produced by the method was left as is.
\item finally, a number of sites can be reduced to their `asymmetric unit coordinates' in order to identify equivalent sites. This is done by taking the symmetry operations for the `pure' structure as a series of rotation matrices and translation vectors $\{\mathbf{R}_i, \mathbf{t}_i\}$. Then for a muon site expressed in fractional coordinates $\mathbf{f}$ we can compute all the periodic images:

\begin{equation}\label{symop}
\mathbf{f}_i = \mathbf{R}_i\mathbf{f}+\mathbf{t}_i
\end{equation}

and retain only one that is closest to a chosen reference point. In this way, all apparent differences due to symmetry operations are removed, and effectively equivalent defects will be squashed in close proximity one to each other, making clustering operations based on position easier. This functionality is provided in Soprano, by the function \texttt{soprano.utils.compute\_asymmetric\_distmat} and makes use of the symmetry analysis library Spglib \cite{spglib}.
\end{enumerate}

Clustering was carried out with the k-means method, using a previous hierarchical step as a way to estimate the number of expected clusters, similarly to what was done in our previous work \cite{Liborio:2018}. After classifying the sites with clustering, the clusterings obtained from CASTEP and DFTB+ results need to be compared. This is done in two ways, using both a distance matrix and a normalized confusion matrix \cite{fowlkes:1983}. Both matrices have the purpose of comparing each possible pair of clusters (one from the CASTEP, the other from the DFTB+ results) with some metric describing their similarity. For the distance matrix we first compute an average muon fractional position for each cluster in both sets, and then compute the distance between all pairs of these centers. This is done while carefully accounting for all symmetries of the crystal to make sure that all such distances are computed between the shortest equivalent sites, using the same asymmetric unit coordinates approach described above. Since these are only fractional coordinate distances, they don't really have a physical meaning, but they correspond to the same criterion with which the clusters were formed, and thus we expect two clusters describing the same stopping site to come out very close with this metric. The closer to zero the distance, the more similar the site represented by the two clusters.\newline
On the other hand, the normalized confusion matrix has the purpose of evaluating how many of the structures, starting from the same point, end up converging to the same cluster through the optimization process. For two clusterings $\mathcal{C}$ and $\mathcal{C}'$, the normalized confusion matrix element $M_{ij}$ is defined as:

 \begin{equation}\label{confmat}
     M_{ij} = \frac{|C_i\cap C_j'|}{\sqrt{n_in'_j}}
 \end{equation}

 where  $|C_i\cap C_j'|$ represents the amount of structures shared by clusters $C_i$ and $C'_j$, which belong to the first and second clusterings respectively, and $n_i$ and $n'_j$ are their respective number of elements. If clusters $C_i$ and $C'_j$ are identical, then $n_i = n'_j = |C_i\cap C_j'|$ and $M_{ij} = 1$.  Conversely, if no structures are shared between $C_i$ and $C'_j$, $M_{ij} = 0$. Therefore, the confusion matrix for two identical clusterings would be square, have only one element set to one per row and column, and every other element would be zero.
 
\section{\label{sec:res}Results and Analysis}

Table \ref{tab:clusters} shows the main features of all the clusters obtained using CASTEP and DFTB$+$ calculations.

\begin{table}[htb]
\begin{tabular}{|C{17mm}|C{22mm}|C{17mm}|C{20mm}|}
\hline
\multicolumn{4}{|c|}{\textbf{CASTEP}} \\
\hline
\textbf{System} & \textbf{Cluster Type} & \textbf{ N$^{\circ}$ of Clusters} & \textbf{Structures per Cluster} \\
\hline
& S1 & 3 & 82,35,8 \\
& C2 & 2 & 14,19 \\
Bithiophene & C3 & 3 & 34,68,1 \\
& C4 & 3 & 3,17,16 \\
& C5 & 2 & 49,35 \\
& float & 3 & 9,1,1 \\
\hline
& N1 & 4 & 57,60,49,50 \\
& C2 & 4 & 11,10,19,23 \\
TCNQ & C3 & 2 & 31,26 \\
& C4 & 2 & 14,10 \\
& C5 & 4 & 30,29,32,36 \\
& float & 2 & 32,1 \\
\hline
& C1 & 3 & 2,1,1 \\
& C2 & 4 & 24,27,19,16 \\
Durene & C3 & 2 & 80,124 \\
& float & 4 & 3,83,65,1 \\
\hline 
\multicolumn{4}{|c|}{\textbf{DFTB$+$}} \\
\hline
\textbf{System} & \textbf{Cluster Type} & \textbf{ N$^{\circ}$ of Clusters} & \textbf{Structures per Cluster} \\
\hline
& S1 & 4 & 31,32,1,1 \\
& C2 & 3 & 13,17,1 \\
& C3 & 3 & 50,69,1 \\
& C4 & 3 & 2,11,15 \\
Bithiophene & C5 & 2 & 43,70 \\
& S(1,7) & 1 & 1 \\
& S(5,6) & 2 & 4,1 \\
& S(1,3) & 3 & 3,1,1 \\
& float & 2 & 2,15 \\
\hline
& N1 & 4 & 57,45,28,46 \\
& C2 & 3 & 7,6,11 \\
& C3 & 3 & 28,39,1 \\
TCNQ & C4 & 3 & 7,5,1 \\
& C5 & 6 & 28,4,34,29,7,24 \\
& float & 2 & 7,1 \\
& other & 1 & 5 \\
\hline 
& C1 & 4 & 4,2,1,1 \\
& C2 & 5 & 29,26,17,23,2 \\
Durene & C3 & 2 & 97,186 \\
& C-CC & 5 & 2,3,1,1,1 \\
& float & 4 & 13,1,7,18 \\
\hline
\end{tabular}
\caption{Clusters obtained from CASTEP and DFTB$+$ calculations on bithiophene, TCNQ and durene muoniated structures, split by molecular site. Each cluster represents a crystallographically inequivalent version of the same molecular site.}
\label{tab:clusters}
\end{table}

The clusters are grouped by `type', meaning whether they're floating sites (float) bonded to some other atom, or anything else (`other'). Bonded sites are labelled by element and site number for those that are recognizable (the numbers are the same as in figures \ref{fig:bithiophene-numbered}, \ref{fig:TCNQ-numbered} and \ref{fig:durene-numbered} respectively for bithiophene, TCNQ and durene) (see Suplementary Information S2 for examples), and by the signature of the function used to identify them when it did not correspond to any recognizable site.
The first important thing to notice from table \ref{tab:clusters} is that all the bonded and float-type clusters predicted by CASTEP were also predicted by DFTB$+$, i.e. there are no false negatives. In the case of bithiophene and durene, DFTB$+$ predicted some extra bonded structures. For bithiophene these are indicated by the distances between the muon and both sulfur atoms in each molecule.  These distances are indicated between brackets as the number of consecutive bonds present between the muon and each sulphur atom in each molecule. The extra bonded structures in bithiophene then are S(1,7), S(5,6) and S(1,3). These linking patterns would not be possible unless some of the molecular bonds were broken - some specific examples are offered in S1 in the Supplementary Information. In general, it seems like DFTB+ has a higher tendency than CASTEP to break bonds in unphysical ways. However, it's easy to dismiss such sites on account of their tiny numbers compared to the other clusters. The C-CC cluster in durene represents a muon bonded to a carbon that is, in turn, bonded only to two more carbons, probably the result of the muon replacing one hydrogen in the aromatic ring, another unlikely reaction which relies on bond breaking. Similarly, the 'other' cluster identified for TCNQ, corresponds to a muon site that is neither bonded nor floating.  This cluster is formed by a small number of structures where the muon has, for instance, connected two TCNQ molecules. All of these examples can be considered as false positives when using DFTB+ and, while they clutter the space of solutions, they don't cause us to lose any real important information.

It becomes important then to compare the CASTEP and DFTB+ clusterings to verify how well they map onto each other. As explained in Section \ref{subs:clust}, we do this in two ways: a distance matrix and a normalised confusion matrix. These two tools are complementary in covering different ways in which the clusters can be considered similar. The distance matrix describes how close two clusterings are to describing the same muon stopping sites; the confusion matrix on the other hand gives us a sense of how much two different clusterings overlap in terms of which structures end up in any given class. Since the structures here are labelled based on their initial configuration (identical for both CASTEP and DFTB+ calculations), the latter gives us an idea of how similar the potential landscape is for the two calculation methods. Fundamentally, if the distance matrix shows a good correspondence, that means the two random searches end up converging to similar minima; but if the confusion matrix shows the same correspondence, that also means that the same starting points end up in the same minima, which is a far stronger sign of overlap between the two methods.\newline

\begin{figure*}[ht]
    \centering
    (a1)\includegraphics[width=0.44\textwidth]{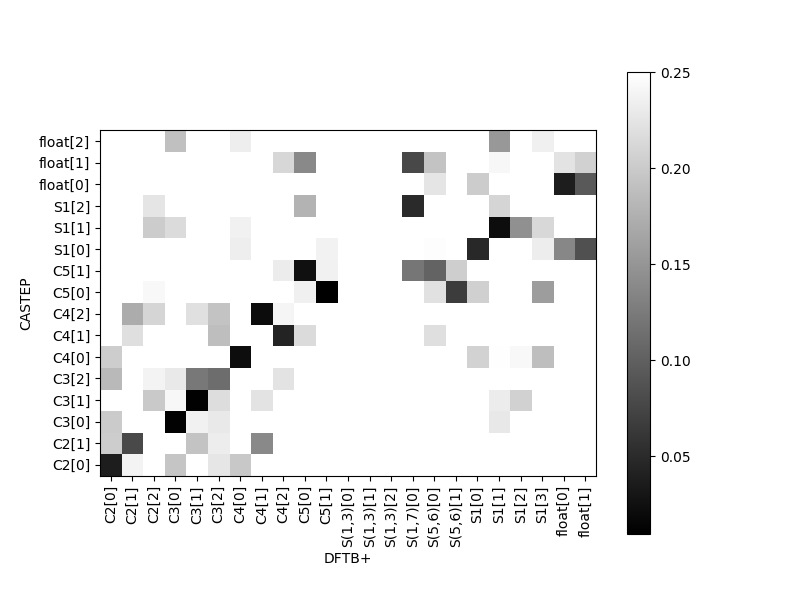}
    (a2)\includegraphics[width=0.44\textwidth]{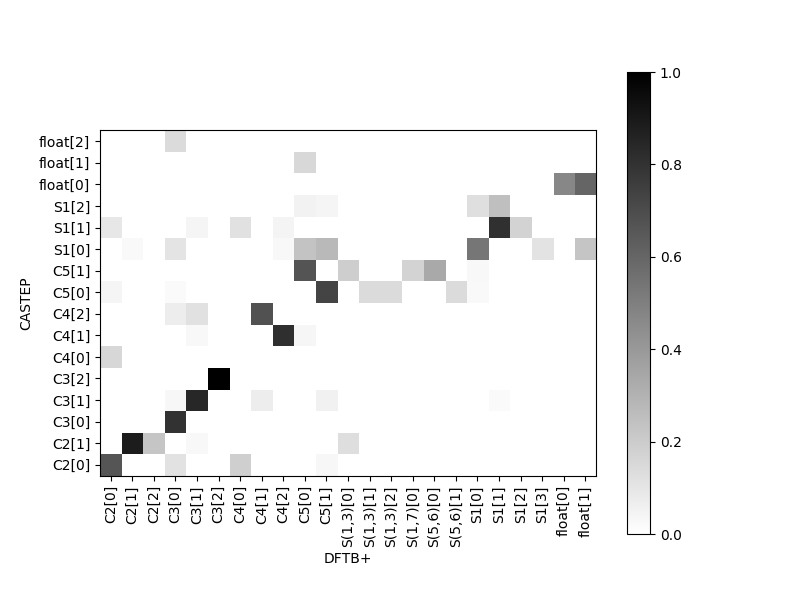} \\
    (b1)\includegraphics[width=0.44\textwidth]{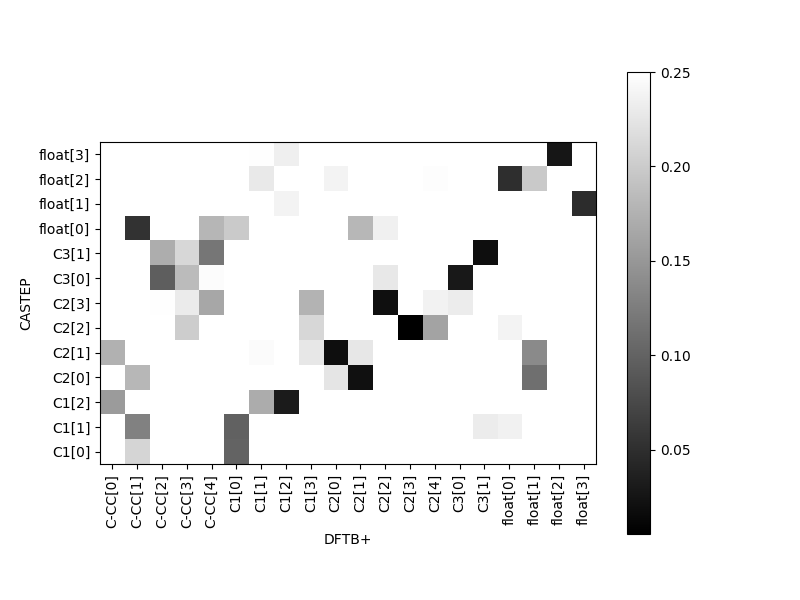}
    (b2)\includegraphics[width=0.44\textwidth]{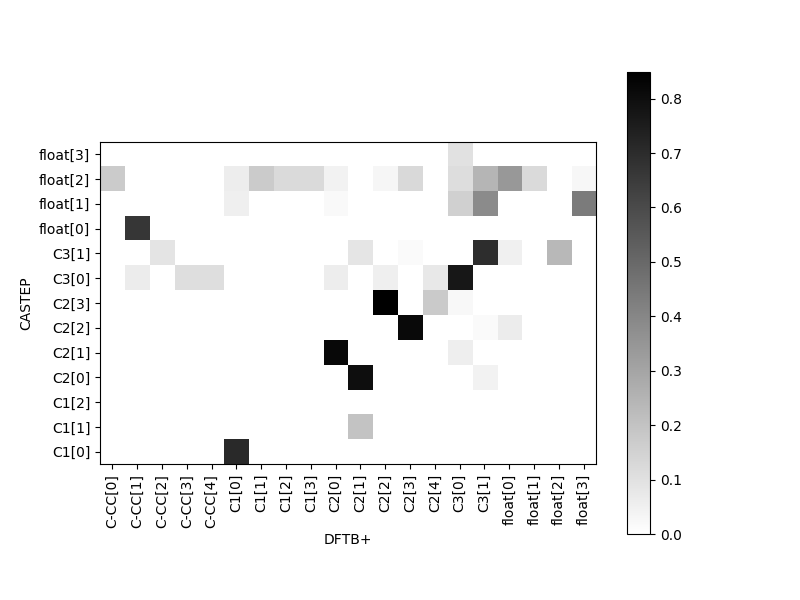} \\
    (c1)\includegraphics[width=0.44\textwidth]{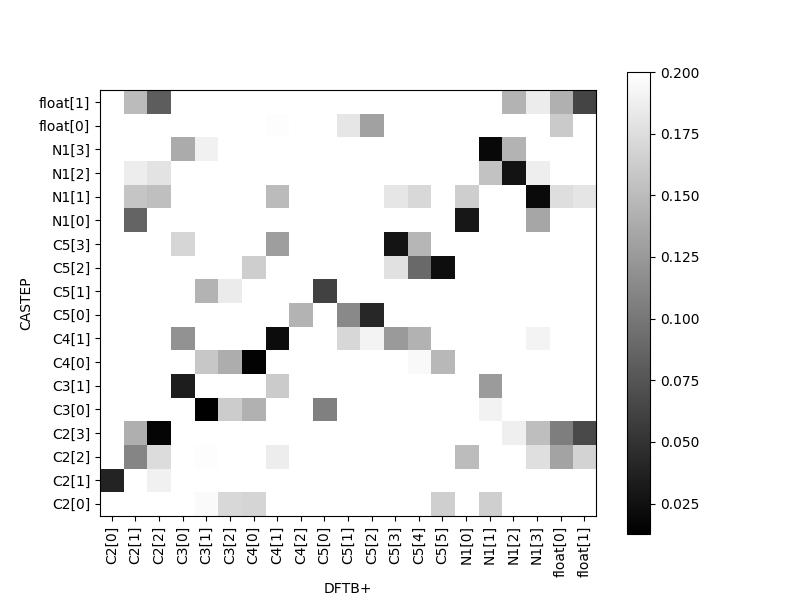}
    (c2)\includegraphics[width=0.44\textwidth]{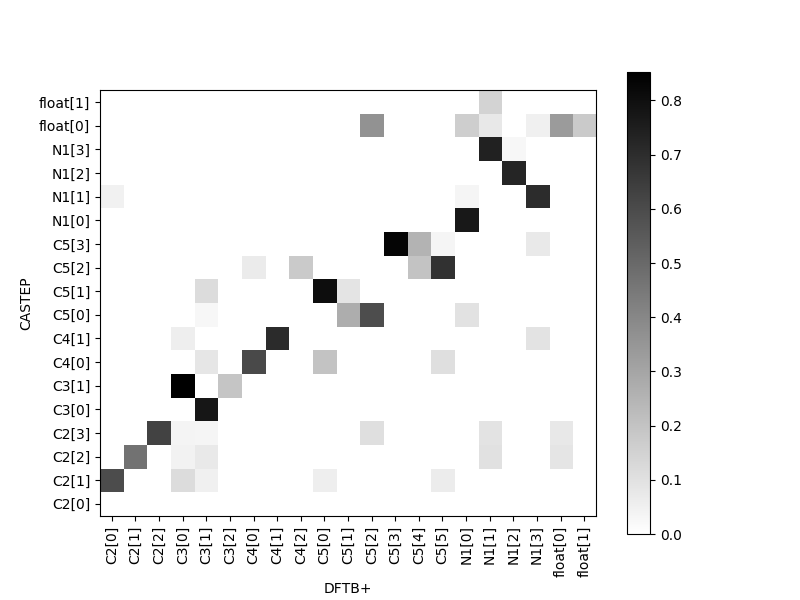} \\
    \caption{Distance (1) and confusion (2) matrices for bithiophene, durene, and TCNQ (a, b and c respectively), plotted with colour scales chosen to highlight the key features. For distances, short values indicate similarity, while for confusion matrices values closer to 1 do. A clear correspondence between the two types of matrix can be seen for all three systems.}
    \label{fig:cluster_matrices}
\end{figure*}

Figure \ref{fig:cluster_matrices} shows the distance and normalised confusion matrices for the clusterings obtained with DFTB+ and CASTEP. The correspondence between clusters obtained with the two methods, for the molecular and floating sites,  can be seen clearly in the form of black squares (representing low distance or high overlap, respectively, for distance and confusion matrices). The sites that tend to lack an assignment are generally the ones that appeared only in DFTB+ (such as the unconventional $S(1,3)$ in bithiophene or the C-CC in durene). The C1 site in durene does the same, and while in this case it appears in CASTEP too, it still requires the unbinding of a hydrogen from a $CH_3$ group to happen, a process that seems unlikely to happen on the time scales of a real muon experiment. Correspondence between distance and confusion matrices, for the molecular and floating sites, is also clearly evident, an excellent indicator of the fact that the potential landscape between CASTEP and DFTB+ is similar. In other words, not only the minima are in similar places, but in general, if the muonium atom is placed in a similar starting spot, it tends to end up in the same one too after geometry optimization.\newline

\begin{figure*}
    (a)\includegraphics[width=0.44\textwidth]{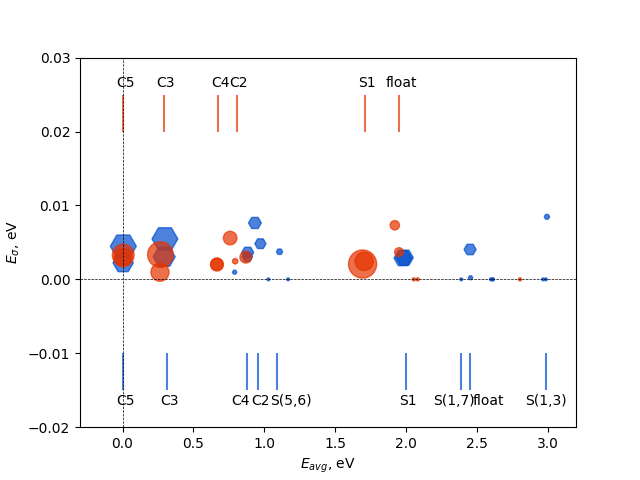} 
    (b)\includegraphics[width=0.44\textwidth]{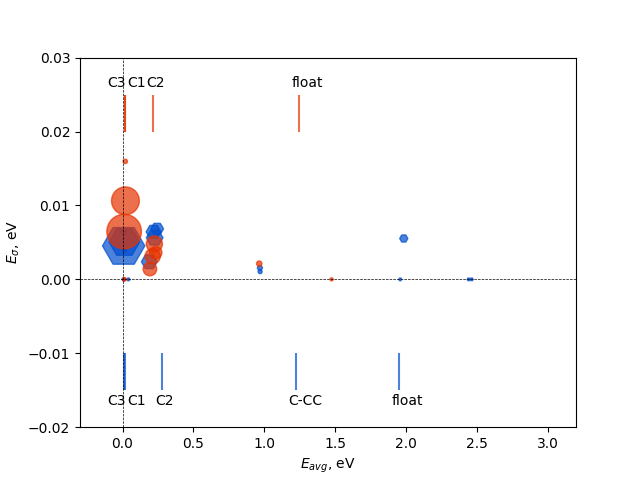} \\
    (c)\includegraphics[width=0.44\textwidth]{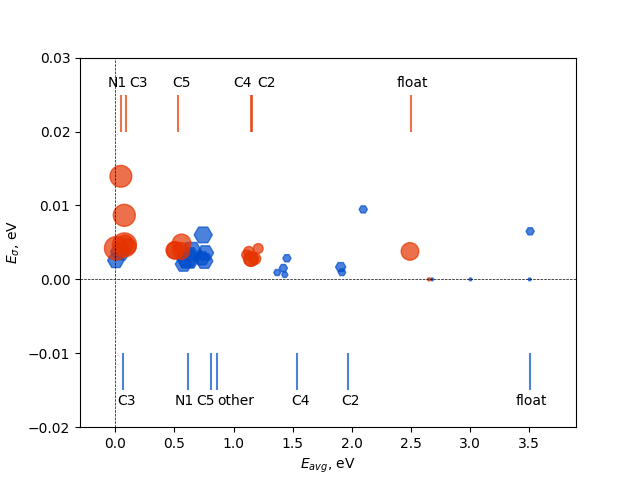}
    \caption{Average-standard deviation plots for energy of clusters for bithiophene (a), durene (b) and TCNQ. The circles represent CASTEP generated clusters, the hexagons are DFTB+. Size of the markers is proportional to the number of structures in each cluster. The average energies for structures classified by molecular sites are marked in each plot by lines for CASTEP-optimized structures (top) and DFTB+ ones (bottom).}
    \label{fig:cluster_eplots}
\end{figure*}

Figure \ref{fig:cluster_eplots} shows the clusters as dots positioned by average and standard deviation of the energies of the structures they contain. Labels are added to show the average energies for structures classified by site for CASTEP and DFTB+ results. Low standard deviations indicate a more `consistent' cluster, with less dispersed energies, thus better convergence to the minimum. Energies are naturally different from different sites, though they tend to vary very little for crystallographically inequivalent realizations of the same molecular site. More importantly, CASTEP and DFTB+ results tend to match remarkably well in energy in a lot of cases. Low energy sites tend to match especially well (e.g. C5 and C3 in bithiophene, C3 and C2 in durene, and C3 in TCNQ). Some of the higher energy sites (S1 in bithiophene, C4 and C2 in TCNQ, and the floating sites in general) have growing discrepancies as the energies get higher. The biggest error appears for the N1 site in TCNQ, which is the lowest in energy in CASTEP but is approximately 0.6 eV higher in DFTB+.

\section{Conclusions}

In this work, the muon stopping sites predicted for bithiophene, TCNQ and durene, by using a DFT-based methodology, were compared with those predicted using a higher level approximation that uses Density Functional Tight Binding (as implemented in the DFTB$+$ code).   

The potential muon stopping sites found by the DFTB$+$-based methodology agree very well with those found by the CASTEP-based methodology.  Moreover, DFTB$+$ calculations are computationally much cheaper than DFT calculations and could potentially be used for treating very large organic systems such as polymers or proteins. This work offers a strong case that it would be reasonable to do so, while also highlighting what are the limitations that a user of the tight binding approach should be watching out for.

\section{Acknowledgements}

The authors are grateful for the
computational support provided by (a) STFC Scientific Computing Department’s SCARF cluster; and (b) the UK Materials and
Molecular Modelling Hub for computational resources, which
is partially funded by EPSRC (No. EP/P020194/1).  Funding for this work was provided by the STFC-ISIS muon source and by the CCP for NMR Crystallography, funded by EPSRC Grant Nos. EP/J010510/1 and EP/M022501/1.

\bibliography{paper}

\end{document}


\maketitle

 
\section{S1 - Identification of Molecular Bonding-sites for the Muon} \label{Bonding-site ID}

Muonic molecular bonding sites were classified by picking a suitable method that would produce a unique identifier for each system. These identifiers were then computed using Soprano (which in turn relied on NetworkX for these analyses) and the structure collections were split making use of them. The identifiers chosen were:

\begin{itemize}
    \item for TCNQ, the ordered pair of shortest network distances (two integers $\geq 1$) from the muon to the two closest nitrogen atoms in the molecule;
    \item for bithiophene, the same, but using the sulphur atoms;
    \item for durene, the alphabetically ordered list of chemical symbols of atoms bonded to the carbon to which the muon was connected (other than the muon itself).
\end{itemize}

It can be seen that these measures are easy to compute once the bonding structure is known and, for well-formed molecules of the given compounds, are unique to each molecular site. Table \ref{tab:bondlabels} contains the values of these identifiers for each of the `regular' molecular sites. When found, these values were interpreted as meaning the muon had bonded to that site.

\begin{table}[]
    \centering
    \begin{tabular}{| c | c | c | c | c | c |}
    \hline
        \multicolumn{6}{|c|}{TCNQ} \\
        \hline
        Site & 1 & 2 & 3 & 4 & 5 \\
        \hline
        Identifier & N(1,5) & N(2,4) & N(3,3) & N(4,4) & N(5,5) \\
        \hline
        \multicolumn{6}{|c|}{Bithiophene} \\
        \hline
        Site & 1 & 2 & 3 & 4 & 5 \\
        \hline
        Identifier & S(1,4) & S(2,3) & S(3,4) & S(3,5) & S(2,5) \\
        \hline
        \multicolumn{6}{|c|}{Durene} \\
        \hline
        Site & 1 & 2 & 3 &  &  \\
        \hline
        Identifier & C-CHH & C-CCC & C-CCH & & \\
        \hline
    \end{tabular}
    \caption{Correspondence between site numbering (as defined in the main paper) and identifying labels used for classification.}
    \label{tab:bondlabels}
\end{table}

\begin{figure}[htp]
  \begin{center}
    \includegraphics[scale=0.6]{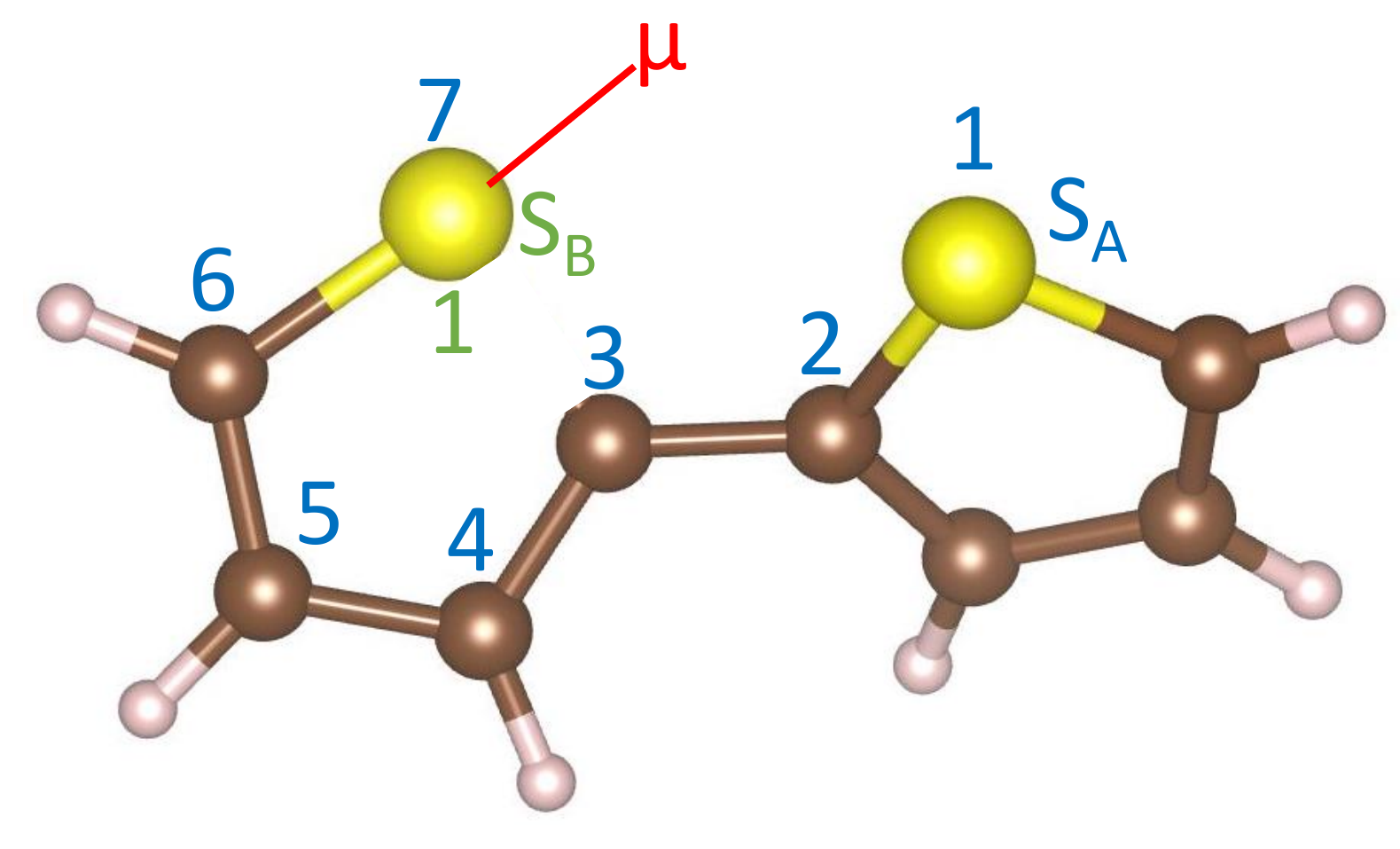} \\
    \end{center}
  \caption{Structure of the S(1,7) muon site, where the broken aromatic ring can be seen. The numbers represent the network distances from the muon site to the sulphur atoms labelled as S$_{A}$ and S$_{B}$ in the molecule. The distances to the two sulphurs are clearly shown with sets of different coloured numbers.}
  \label{muon-location}
\end{figure}

In some cases, the muon connected to the molecule in a way that did not fit within the categories defined in Table \ref{tab:bondlabels}, for example by breaking a bond. This has happened mainly in DFTB+ calculations. In this case, the identifiers have been used as they were.
Figure (\ref{muon-location}) shows an example of such a situation. Here the muon has broken one of the C-S bonds, leading to an anomalous S(1,7) label. 

\section{S2 - Location of Representative Stopping Sites}  \label{representative stopping sites}

\begin{figure}[htb]
(a)\includegraphics[width=0.34\textwidth]{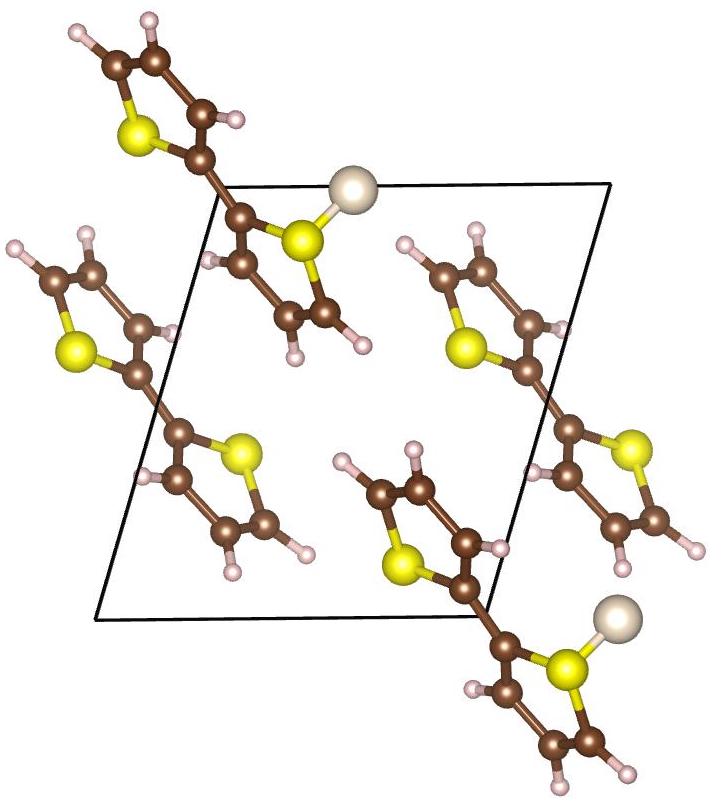} 
(b)\includegraphics[width=0.34\textwidth]{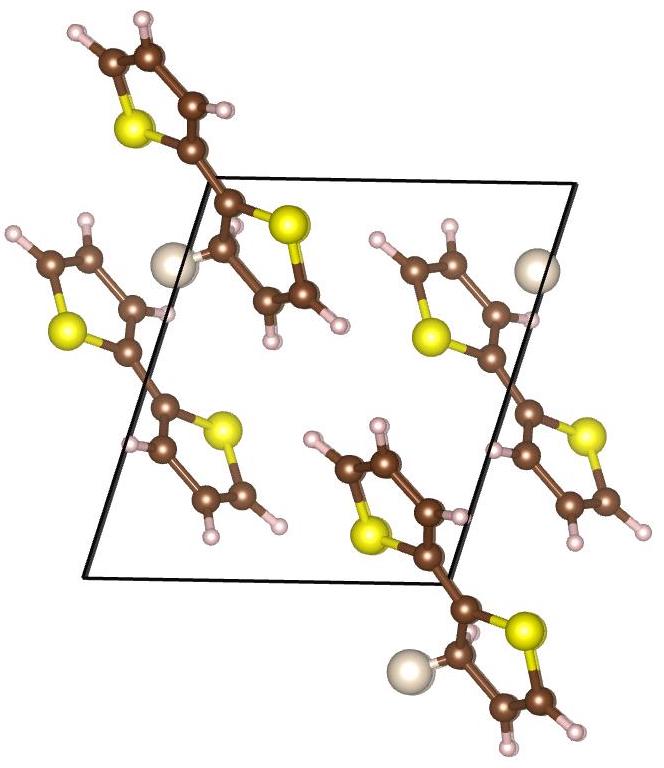} \\
(c)\includegraphics[width=0.47\textwidth]{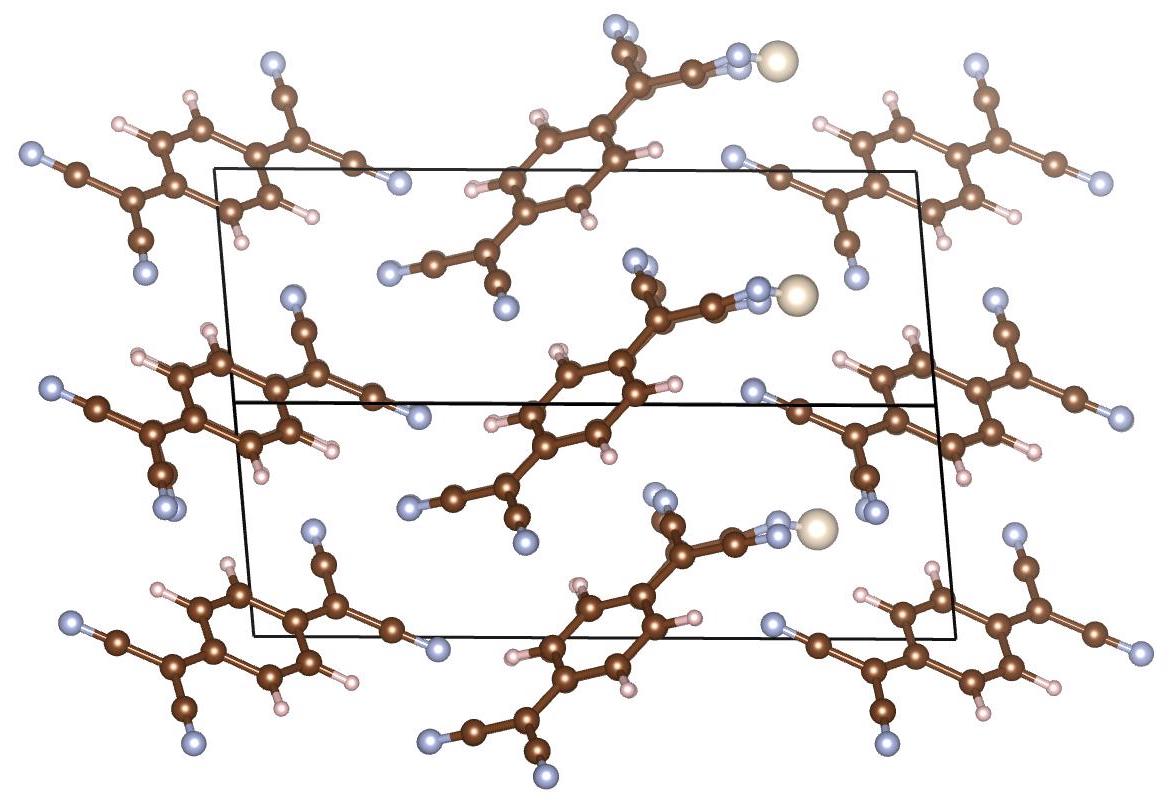} 
(d)\includegraphics[width=0.46\textwidth]{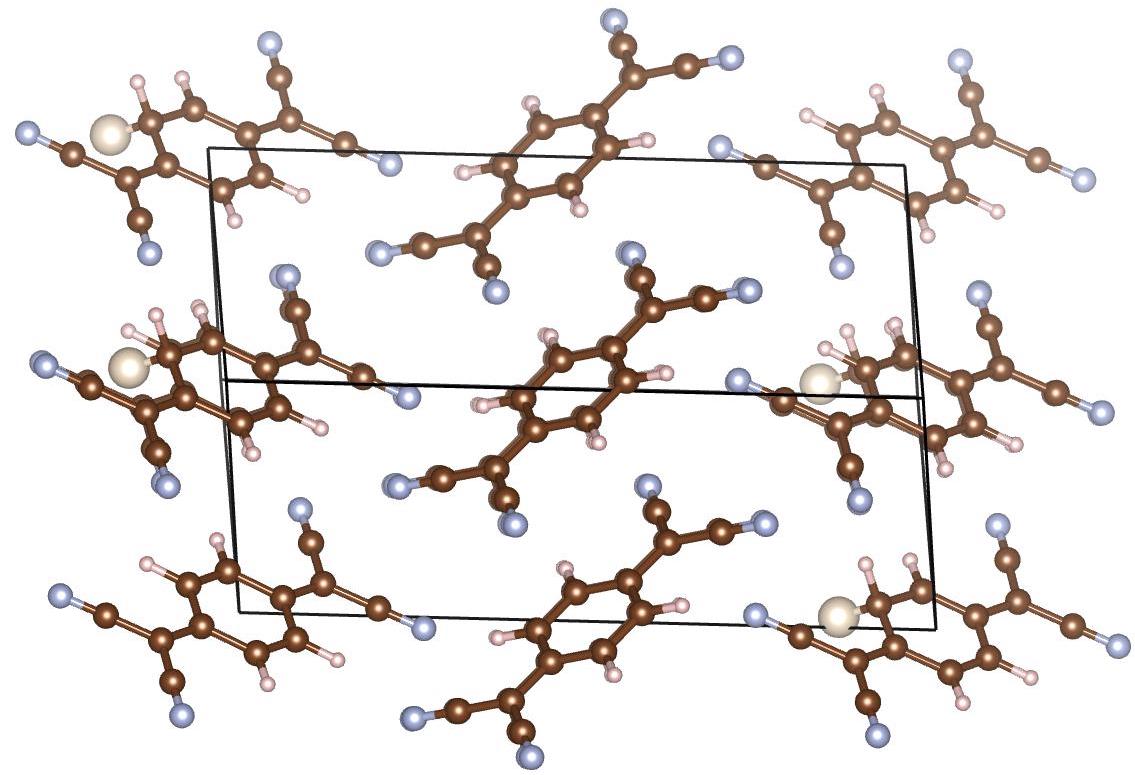} \\
(e)\includegraphics[width=0.45\textwidth]{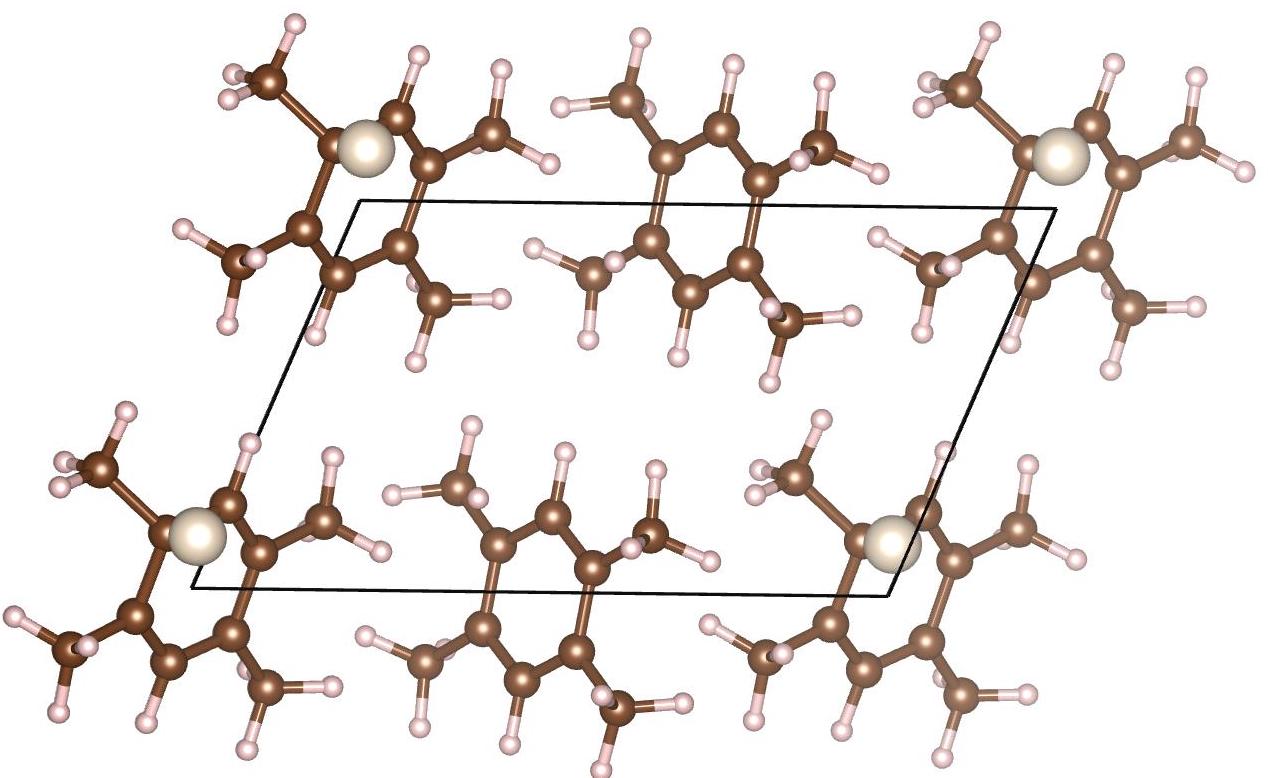}
(f)\includegraphics[width=0.44\textwidth]{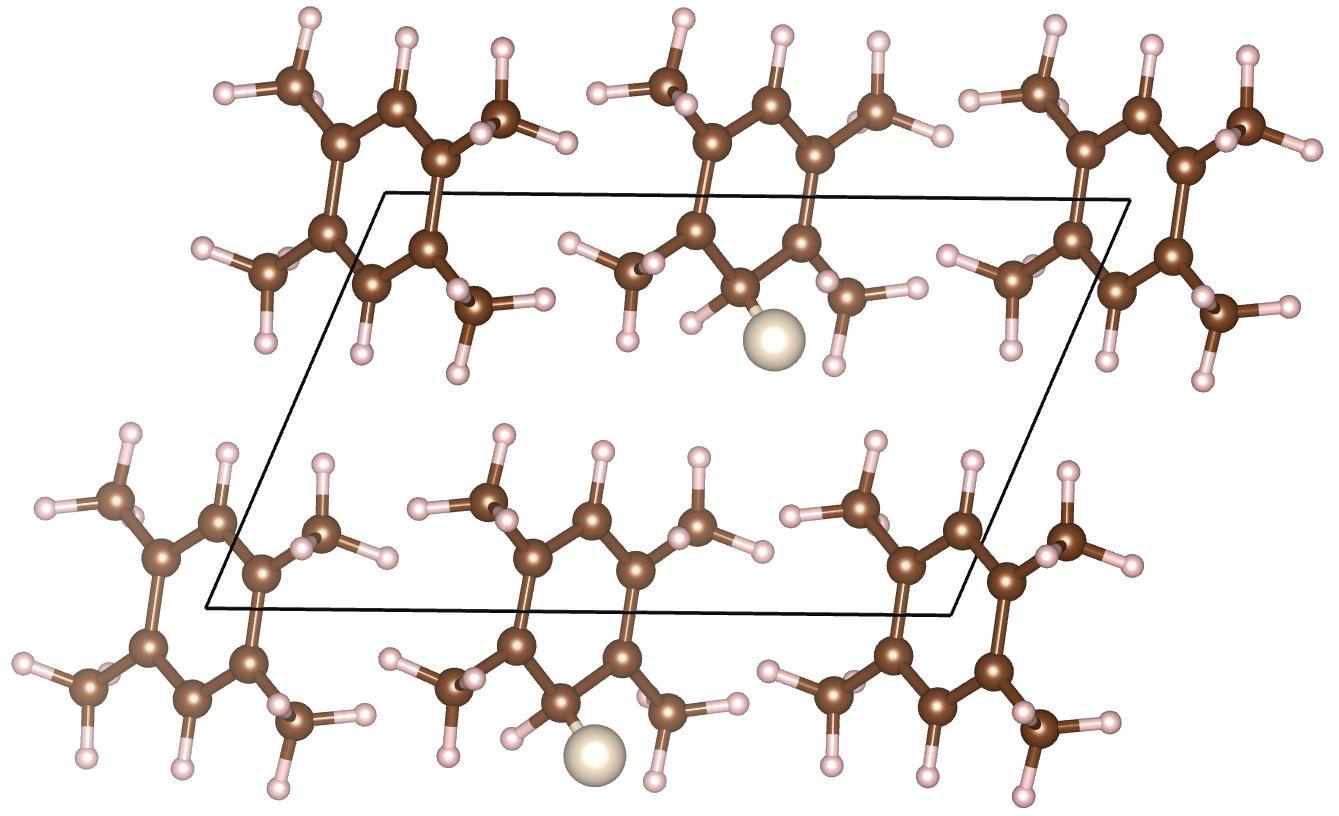} 
\caption{The muon is indicated as a large white atom.
Figures (a) and (b) show bithiophene with the muon bonded to one of the sulphur atoms -S1- and to one of the carbon atoms in the aromatic ring -C3-. Figures (c) and (d) refer to TCNQ and show the muon attached to a nitrogen atom -N1- and to one of the carbon atoms in the aromatic ring -C5-. Finally, figures (e) and (f) show the two main bonding sites in durene, which are both in carbon atoms in the aromatic ring: C2 and C3.}
\label{fig:RepreSites}
\end{figure}

Figure \ref{fig:RepreSites} shows examples of representative stopping bonded sites for muonium in the three crystalline compounds.  These sites are typical of aromatic compounds, where the muon attacks regions of high electronic densities, such as double and triple bonds, breaks one of these bonds and release an unpaired electron. 

